\documentclass[preprint,showpacs,preprintnumbers,amsmath,amssymb,nofootinbib]{revtex4}

\usepackage{etex}

\usepackage{amsthm,amscd,amsbsy,array}
\usepackage{bm}% bold math
\usepackage{soul} % underline, strikethrough, etc.

\usepackage{graphics,graphicx,xcolor}

\usepackage{soul} % underline, strikethrough, 

\usepackage[colorlinks=true, pdfstartview=FitV, linkcolor=blue, citecolor=blue, urlcolor=blue]{hyperref} % hyperref

\newcommand{\red}{\textcolor{red}}  
\newcommand{\blue}{\textcolor{blue}}

\newcommand{\gb}{\colorbox{green}}

\newcommand{\dgreen}{\textcolor[rgb]{0,0.35,0}}

\newenvironment{redtext}{\color{red}}{\ignorespacesafterend} 
\newenvironment{bluetext}{\color{blue}}{\ignorespacesafterend}

\newcommand{\bblue}{\begin{bluetext}} 
\newcommand{\eblue}{\end{bluetext}} 
\newcommand{\bred}{\begin{redtext}}
\newcommand{\ered}{\end{redtext}}

\numberwithin{equation}{section}

\let\ssection=\section
\renewcommand{\section}{\setcounter{equation}{0}\ssection}

\newcommand{\bA}{{\bf A}}
\newcommand{\cA}{{\mathcal{A}}}

\newcommand{\bb}{{\bf b}}
\newcommand{\bB}{{\mathbf{B}}}

\newcommand{\bbeta}{\boldsymbol{\beta}}

\newcommand{\bone}{\boldsymbol{1}}

\newcommand{\bc}{{\mathbf{c}}}
\newcommand{\bC}{{\mathbf{C}}}

\newcommand{\diag}{\mathrm{diag}}

\newcommand{\da}{\dot{a}}

\newcommand{\db}{\dot{b}}

\newcommand{\dgamma}{\dot{\gamma}}

\newcommand{\ddchi}{\ddot{\chi}}

\newcommand{\rg}{\mathrm{g}}

\newcommand{\bgamma}{\boldsymbol{\gamma}}

\newcommand{\bk}{\mathbf{k}}
\newcommand{\bK}{{\bf K}}

\newcommand{\bp}{{\bf p}}

\newcommand{\bx}{{\bm{x}}}

\newcommand{\bbR}{\mathbb{R}}

\newcommand{\Tr}{\mathrm{Tr}}

\newcommand{\bX}{{\bm{X}}}
\newcommand{\bY}{{\bm{Y}}}

\newcommand{\by}{{\bf y}}

\def\smallover#1/#2{\hbox{$\textstyle\frac{#1}{#2}$}} %
\def\Rarrow{\quad\Rightarrow\quad}

\def\bp{{\bm{p}}}

\def\and{{\qquad\text{and}\qquad}}
\def\benu{\begin{enumerate}}
\def\eenu{\end{enumerate}}
\def\beq{\begin{equation}}
\def\eeq{\end{equation}}
\def\beqa{\begin{eqnarray}}
\def\eeqa{\end{eqnarray}}

\def\barray{\left(\begin{array}}
\def\earray{\end{array}\right)}
\def\barraynb{\begin{array}}
\def\earraynb{\end{array}}

\def\?{\quad{\gb{\fbox{\texttt{?}}\;}}\quad}
\def\p{{\partial}}

\def\v0{\mathbf{0}}

\def\6{\partial}
\def\7{\tilde}
\def\8{\widehat}
 \def\bx{{\bf x}}

\newcommand{\hbx}{{\hat{\mathbf{x}}}}
\newcommand{\hu}{{\hat{u}}}
\newcommand{\hv}{{\hat{v}}}

\newcommand{\const}{\mathop{\rm const.}\nolimits}
\newcommand{\half }{\frac{1}{2}}

\def\smallover#1/#2{\hbox{$\textstyle\frac{#1}{#2}$}} %
\def\smallcirc{{\raise 0.5pt \hbox{$\scriptstyle\circ$}}}
\def\2{{\smallover1/2}}

\newcommand{\medbox}[1]{\fbox{%
\rule[-10pt]{0pt}{25pt}$\;\;\displaystyle{#1}\;\;$}%
}

\let\ssection=\section
\renewcommand{\section}{\setcounter{equation}{0}\ssection}

\begin{document} % ££

\preprint{arXiv: 1709.02299v3 [gr-qc]}

\title{Memory Effect for Impulsive Gravitational Waves}

\author{
P.-M. Zhang${}^{1}$\footnote{e-mail:zhpm@impcas.ac.cn},
C. Duval${}^{2}$\footnote{
%Aix-Marseille Universit\'e, CNRS, CPT, UMR 7332, 13288 Marseille, France.
%Universit\'e de Toulon, CNRS, CPT, UMR 7332, 83957 La Garde, France.
mailto:duval@cpt.univ-mrs.fr},
P. A. Horvathy${}^{1,3}$\footnote{mailto:horvathy@lmpt.univ-tours.fr},
}

\affiliation{
$^1$Institute of Modern Physics, Chinese Academy of Sciences, Lanzhou, China
\\
$^2$Aix Marseille Univ, Universit\'e de Toulon, CNRS, CPT, Marseille, France
\\
$^3$Laboratoire de Math\'ematiques et de Physique
Th\'eorique,
Universit\'e de Tours,
France
}

\date{Jan 23, 2018}


\begin{abstract}
Impulsive gravitational plane waves, which have a $\delta$-function singularity on a hypersurface, can be obtained by squeezing smooth plane gravitational waves with Gaussian profile. They exhibit (as  do their smooth counterparts) the Velocity Memory Effect: after the wave has passed, particles initially at rest move apart with non vanishing constant transverse velocity. 
A new effect is that, unlike to the smooth case, (i) the  velocities of particles originally at rest jump, (ii) the spacetime trajectories become discontinuous along the (lightlike) propagation direction of the wave.
%\\
%{\bf Shock -- V3}
\end{abstract}

\maketitle

\tableofcontents

%%%%%%%%%%%%%%%%%%%%%%%%%%%%%%%%%%%%%%%%%%%%%%%%%%%%%%%%%%%%%%%%%%%%%%%%%%%%%%
%%%%%%%%%%%%%%%%%%%%%%%%%%%%%%%%%%%%%%%%%%%%%%%%%%%%%%%%%%%%%%%%%%%%%%%%%%%%%%
\section{Introduction}\label{Intro}
%%%%%%%%%%%%%%%%%%%%%%%%%%%%%%%%%%%%%%%%%%%%%%%%%%%%%%%%%%%%%%%%%%%%%%%%%%%%%%
%%%%%%%%%%%%%%%%%%%%%%%%%%%%%%%%%%%%%%%%%%%%%%%%%%%%%%%%%%%%%%%%%%%%%%%%%%%%%%

The displacement of freely falling particles by a gravitational wave,  called the ``memory effect'' 
\cite{ZelPol,BraGri,BraTho,GriPol,BoPi89,Christo,Thor,BlaDam,Fava,Harte,Garfinkle,ShortMemory,LongMemory,Lasenby} has attracted considerable recent attention, 
due to its potential use for detecting gravitational waves \cite{GiHaw,Lasky,Lasenby}. 
%The original prediction of Zel'dovich and Polnarev \cite{ZelPol} claimed that particles at relative rest before the wave arrives would suffer a \emph{permanent  
%displacement with vanishing relative velocity}. This is also what is said for the non-linear generalizations \cite{Christo,Thor,BlaDam,Garfinkle}.

In our previous papers \cite{ShortMemory,LongMemory} we studied  linearly polarized  exact plane waves with a smooth profile and  
 found that after the wave has passed
our particles  fly apart with \emph{constant but non-vanishing velocity}, consistently  with suggestions by Braginsky, Grishchuk, Thorne, and Polnarev \cite{BraGri,BraTho,GriPol}, and by Bondi and Pirani \cite{BoPi89}~: instead of a permanent  displacement, there will be a \emph{velocity memory effect}.

In this Paper we extend our investigations and derive similar results for impulsive waves \cite{%Synge57,Papa59,Treder62,Daut64,Daut69,
Penr2,Sou73,Bini,Griffiths,BarHog,Steinbauer,KunStein,Luk}, which has the advantage that explicit calculations are possible. Our new results are consistent with those  obtained in the smooth case in \cite{ShortMemory,LongMemory}. The novelty is \emph{the velocity jump suffered when the $\delta$-function profile wave passes,  and a discontinuity of the trajectories in the forward direction}.

Our paper is organized as follows. After recalling the three main coordinate systems we use, we study what happens when a Gaussian profile is shrunk to a Dirac $\delta$-function. 
This corresponds to obtaining an impulsive wave  by suppressing the inside zone.
Carroll symmetry \cite{Sou73,Carroll4GW}, outlined in sect. \ref{Carrollsec}, plays a  distinguished r\^ole.
Our main section, \ref{GeoSec}, discusses the geodesics in impulsive gravitational waves in various coordinate systems, followed by a numerical study for Gaussian profile.
Sect. \ref{SteinbauerSec} explains the relation to previous work on impulsive waves \cite{Steinbauer,KunStein}.

\goodbreak
%%%%%%%%%%%%%%%%%%%%%%%%%%%%%%%%%%
\section{Plane gravitational waves}\label{planewavesec}
%%%%%%%%%%%%%%%%%%%%%%%%%%%%%%%%%%%

%%%%%%%%%%%%%%%%%%%%%%%%%%
\subsection{Brinkmann (B) coordinates}
%%%%%%%%%%%%%%%%%%%%%%%%%%

Plane gravitational waves are often described  in 
\emph{Brinkmann coordinates} (B) \cite{Bri} in  terms of which the metric is
\beq
\rg =\delta_{ij}\,dX^idX^j+2dUdV+K_{ij}(U){X^i}{X^j}\,dU^2,
\label{Bplanewave}
\eeq
where the symmetric and traceless $2\times2$ matrix
$K(U)=\left(K_{ij}(U)\right)$ characterizes the profile of the wave.  
In this paper we  consider linearly polarized `` $+$ type" waves with 
\beq
K(U)=
\half\cA(U)\,\diag(1,-1),
\label{polBrink}
\eeq
where $\cA(U)$ is an arbitrary function \footnote{The most general profile is 
\beq
K_{ij}(U){X^i}{X^j}=
\half{\cA_{+}}(U)\Big((X^1)^2-(X^2)^2\Big)+\cA_{\times}(U)\,X^1X^2,
\label{genBrink}
\eeq
where $\cA_{+}$ and $\cA_{\times}$ are the amplitudes of the $+$ and $\times$ polarization states. Although in this paper we focus our investigation at the diagonal case $\cA_{\times}=0$, we prefer to keep our   general formulae in view of later applications  to primordial gravitational waves and CMB (Cosmic Microwave Background).}.

The Brinkmann coordinates $(X^1,X^2,U,V)$ are global, and the transverse spatial distance is simply $|\bX-\bY|=\sqrt{(\bX-\bY)^2}$.

\goodbreak 
%%%%%%%%%%%%%%%%%%%%%%%%%%
\subsection{Baldwin-Jeffery-Rosen (BJR) coordinates}
%%%%%%%%%%%%%%%%%%%%%%%%%%

Another useful description  is provided by \emph{Baldwin-Jeffery-Rosen coordinates} (BJR)  \cite{Sou73,LaLi,BaJeRo}, for which
\beq
\rg=a_{ij}(u)\,dx^idx^j+2du\,dv,
\label{BJRmetrics}
\eeq
with $a(u)=(a_{ij}(u))$ a positive definite $2\times2$ matrix, which is an otherwise arbitrary function of ``non-relativistic time'', $u$ \footnote{Our terminology comes from the ``Eisenhart-Bargmann''  framework \cite{Eisenhart,Bargmann} where $u=U$ becomes indeed non-relativistic time and $v$ resp. $V$ are referred to as ``vertical coordinates''.}. The BJR coordinates $(x^\mu)=(x^1,x^2,u,v)$ are typically not global and suffer from  singularities \cite{Sou73,BoPi89,ShortMemory,LongMemory}.
In these coordinates, the transverse spatial distance involves also the transverse metric, 
$\Vert\bx-\by\Vert=\sqrt{a_{ij}(u)\,(x^i-y^i)(x^j-y^j)}$.

Calling $P(u)$ a square-root of $a(u)$, 
\begin{equation}
a(u)=P(u)^{T}{}P(u),
\label{a=PTP}
\end{equation}
the relation between the two coordinate systems is given by \cite{Gibb75,Carroll4GW} 
\beq
\medbox{
{\bX} =P(u)\,\bx,
\quad
U=u,
\quad
V=v-\frac{1}{4}\bx\cdot\dot{a}(u)\bx,
}
\label{BBJRtrans}
\eeq
where the $2\times2$ matrix $P(u)$ is a solution of the matrix Sturm-Liouville (SL) equations
\beq
\medbox{
\ddot{P}=K\,P
\qquad
\&
\qquad
P^{T}\dot{P}-\dot{P^{T}}P=0.
}
\label{KSL}
\eeq
Here $\dot{P}=dP/du$, and the superscript $T$ denotes transposition. In what follows, we shall agree that $U=u$  denote the same (``non-relativistic time'') coordinate and use one or the other notation just to emphasize which coordinate system we are working with. 
The profiles in the two coordinate systems are related by,
\beq
K= \half P \left(\dot{b} + \half b^2\right) P^{-1}
\qquad
\text{where}
\qquad
b=a^{-1}\dot{a}.
\label{Einstein}
\eeq

Waves  whose profile  vanishes outside an interval $U_i\leq U \leq U_f$ of ``non-relativistic time'', $U$ are called \emph{sandwich waves}.
The regions  $U< U_i,\, U_i \leq U \leq U_f \,,  U_f < U$ are referred to as the \textit{before, inside}, and \textit{after - zones}, respectively \cite{BoPi89}.  The before and after-zones are flat; the inside-zone  is only Ricci-flat, which requires $K$ to be traceless. By (\ref{Einstein})  this amounts to
\beq
\Tr\left(\db+\half{}b^2\right)=0.
\label{Ricci=0}
\eeq
Putting now
$ 
\chi=\big(\det{a}\big)^{\frac{1}{4}}>0
$
and
$
\gamma=\chi^{-2}a,
$ 
eqn. (\ref{Ricci=0}) leads to another Sturm-Liouville equation,
\begin{equation}
\ddchi+\frac{1}{8}\Tr\left((\gamma^{-1}\dgamma)^2\right)\chi=0,
\label{SLchi}
\end{equation}
which guarantees that the vacuum Einstein equations are satisfied for an (otherwise arbitrary) choice of the unimodular symmetric $2\times2$ matrix $\gamma(u)$. The BJR coordinate system is regular as long as $\chi\neq0$.

\goodbreak 

%%%%%%%%%%%%%%%%%%%%%%%%
%\section{Impulsive waves}\label{Impulsivesec}
%%%%%%%%%%%%%%%%%%%%%%%%
 
In this paper, we focus our attention at  \emph{impulsive waves}, which have received extensive attention both from the physical and the mathematical \cite{Penr2,Sou73,Bini,Griffiths,BarHog} and in particular from distribution-theoretical \cite{Steinbauer,KunStein} points of view. Following Penrose, an impulsive gravitational wave is a gravitational wave whose metric is continuous but not $C^1$ on some (null) hypersurface. Its curvature tensor contains therefore a delta-function \cite{Penr2} \footnote{Impulsive waves  should be distinguished from shock waves for which the second derivative of the metric suffers a discontinuity across a (null) hypersurface.}. 
Impulsive waves are sandwich waves whose inside-zone has been suppressed, 
\beq
    U_i=U_f=0.
\label{impuls}
\eeq
The metric is flat both in the before and the after-zones, $U<0$ and $U>0$, respectively; their $\delta$-function behavior is on the hypersurface $U=0$.

\begin{figure}[h]
\includegraphics[width=0.47\textwidth]
{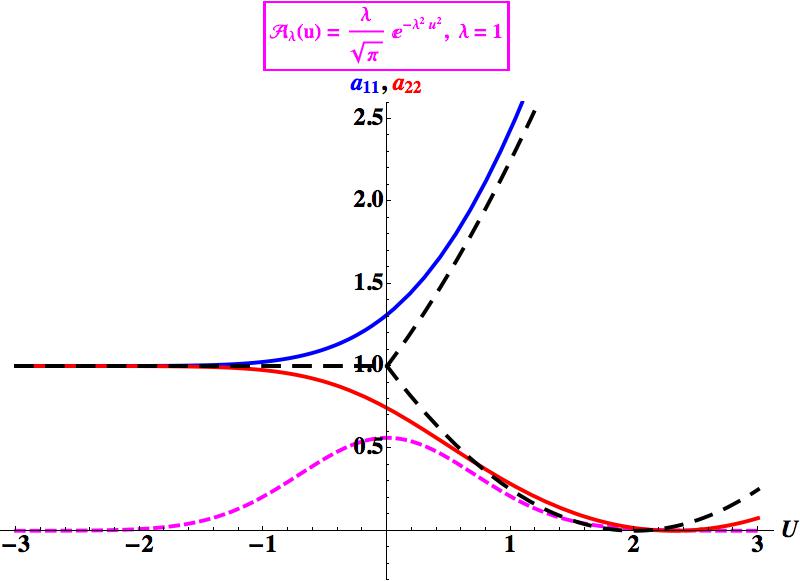}\;
\includegraphics[width=0.47\textwidth]
{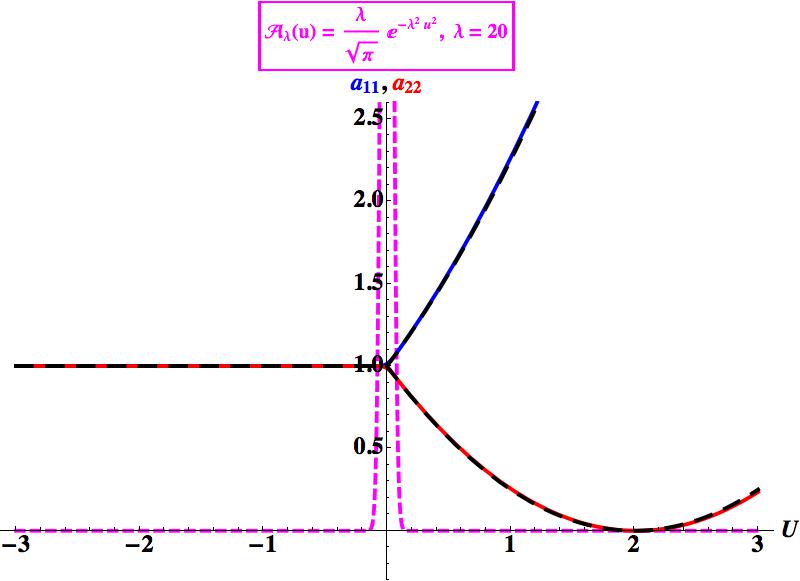}\\
\vskip-8mm
\null\hskip-6mm(a)\hskip80mm(b)\\
\vskip-2mm
\caption{\textit{ 
Squeezing the Gaussians \textcolor{purple}{${\cA}_{\lambda}$} to a Dirac $\delta$-function, the transverse metrics $a_\lambda(u)$ (in \red{\bf red} and \blue{\bf blue}) tend to that of the impulsive wave, (\ref{abeforeafter}) in BJR coordinates, depicted in {\bf dashed black} lines.}}
\label{adeltalimit}
\end{figure}

Flat metrics can be determined explicitly \cite{Sou73,LongMemory}. Using BJR coordinates we assume that the before-zone $u<0$ is described by  inertial coordinates and thus $a_{ij}=\delta_{ij}$.
For $u>0$ the metric is described in turn by a continuous but not necessarily smooth matrix $a_{ij}(u)$.  
  Defining $c_0$ as the right-hand limit of the ``time'' derivative of the transverse metric in the afterzone
$u>0$,  
\beq
c_0=\half\dot{a}(0+),
\label{c0}
\eeq  
and solving the flatness equation $R_{uiuj}=0$ if the metric is given by $a_{ij}=\delta_{ij}$ in the before-zone, the general formul{\ae}\ in~\cite{Sou73,LongMemory} yield,
\begin{equation}
a(u)=\left\{\barraynb{cll}
\bone
\;
&\text{\small for}&u\leq0,
\\ 
(\bone+u\,c_0)^2
\;
&\text{\small for}&u>0.
\earraynb\right.
\label{abeforeafter}
\end{equation}
The symmetric $2\times2$ matrix $c_0$ in (\ref{c0}) characterizes the flat after-zone. Calling $k,\ell$ its (real) eigenvalues we easily find $\chi(u)=\sqrt{(1+u k)(1+u\ell)}$.  Since the SL equation (\ref{SLchi}) holds both for $u\leq0$ and $u>0$, the function $\dot\chi$ is necessarily continuous at $u=0$, hence $\dot{\chi}(0)=0$ because $\dot{\chi}(u)=0$ for all $u\leq0$. Therefore we have $\dot{\chi}(0)=(k+\ell)/(2\chi(0))=0$, implying that $c_0$ has two opposite  eigenvalues $\pm k$ \cite{Sou73}.  In an eigenbasis (which means in fact polarization), we have therefore 
\beq  
c_0=k\,\diag (1,-1)
\label{c00}
\eeq
at $u_0=0+$. We will assume $k\geq0$ with no loss of generality.
If  $k=0$, we would have $a(u)=\bone$ for all $u$ 
and there would be no wave. Henceforth we assume that $k\neq0$ in the after-zone. 
The profile is shown on FIG.\ref{adeltalimit}  is indeed continuous but  non-differentiable at~$u=0$,
\beq 
\dot{a}(u) = 2 (1+u\, c_0) c_0\, \theta(u),
\label{adot}
\eeq 
where $\theta(u)$ is the Heaviside step-function (verifying 
$\theta(0)=0$ and $\theta(0+)=1$).  

%%%%%%%%%%%%%%%
%%%%%%%%%%%%%%%%

The $C^0$ metric (\ref{BJRmetrics}) defined by eqn (\ref{abeforeafter}) describes therefore a plane \emph{impulsive gravitational wave} in the sense of Penrose \cite{Penr2}.

\begin{figure}[h]
\includegraphics[width=0.47\textwidth]{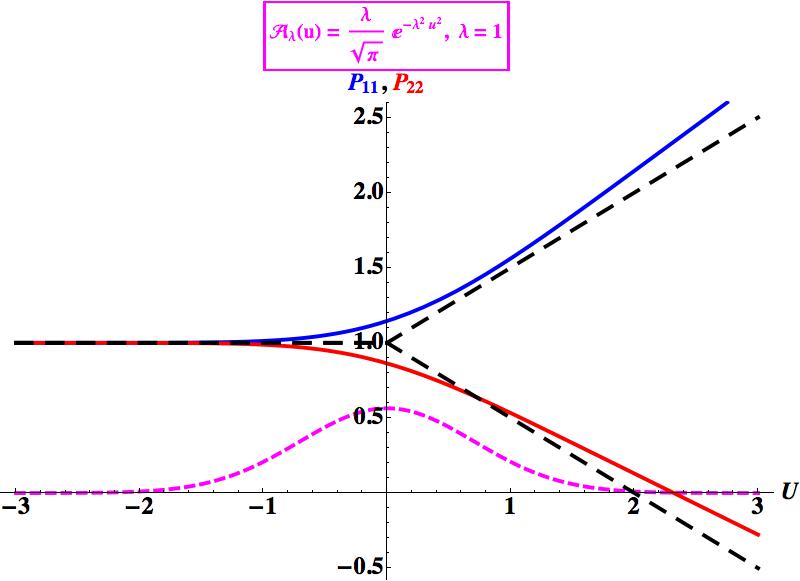}\;
\includegraphics[width=0.47\textwidth]{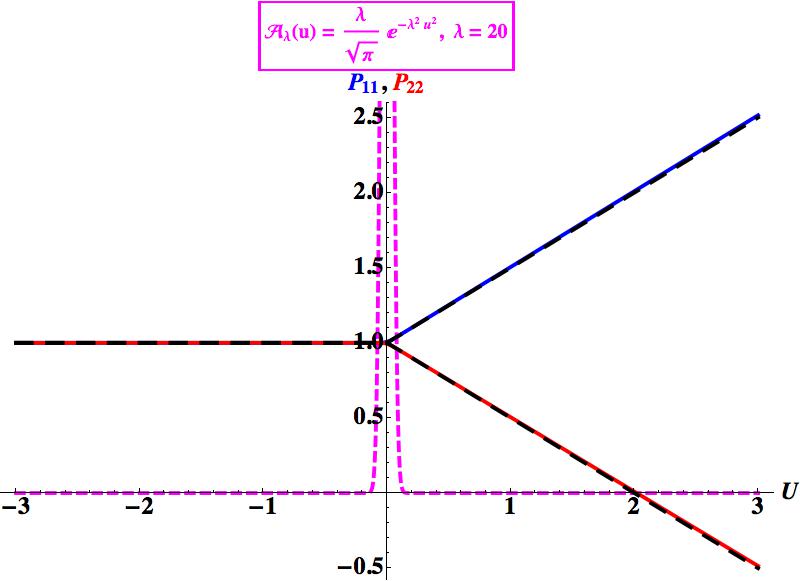}
\\
\vskip-8mm
\null%\hskip1mm
(a)\hskip75mm(b)\\
\vskip-2mm
\caption{\textit{Plotting the numerical solution of the Sturm-Liouville equation (\ref{KSL}) for the profile ${\cA}_{\lambda}(U)=(\lambda/\sqrt{\pi})\,e^{-\lambda^2U^2}$ shows that the components of the diagonal matrix $P_{\lambda}(U)$ approach, for large $\lambda$, those of the impulsive wave (\ref{Pbeforeafter}) [in {\bf dashed black}].}}
\label{Pdeltalimit}
\end{figure}

The transverse matrix $a(u)$ in  (\ref{abeforeafter}) is quadratic in $u$, and  we find actually more convenient to use a symmetric square-root $P(u)$ of $a(u)$, namely $a(u)=P(u)^2$, where
\beq
\medbox{
 P(u)= \bone+u\,\theta(u)\,c_0
 }
\label{Pbeforeafter}
\eeq
is affine in $u$. Its components are shown in Fig.\ref{Pdeltalimit} by {\bf dashed black} lines.

%%%%%%%%%%%%%%%
%%%%%%%%%%%%%%%%
 We note that $\det{P}$  vanishes exactly once, namely at 
$  
u_1=k^{-1}>0,
$ 
signalling that one (but not the other) component of $P(u)$ has a single zero. Realizing that $\det P=0$ iff
 $\chi=0$ \cite{ShortMemory} indicates that
the BJR coordinate system is regular when $u < k^{-1}$.

Turning to  Brinkmann coordinates the  profile, obtained
by substituting (\ref{abeforeafter}) and (\ref{Pbeforeafter})  in (\ref{Einstein}) is \footnote{Our eqns (\ref{abeforeafter})  and (\ref{shockK}) 
correct a sign error in eqns (1) and (2) of Steinbauer et al. \cite{Steinbauer}, cf. \cite{KunStein}.
Moreover, substituting (\ref{Pbeforeafter}) and  (\ref{adot}) in (\ref{BBJRtrans}) yields eqn (3) of \cite{Steinbauer}. eqn (\ref{abeforeafter}) is also consistent with Case 1 of Bini \cite{Bini}.}
\beq
\cA(U) 
=2k\, \delta(U)\, 
\quad\Longleftrightarrow\quad 
c_0=k\,\diag (1,-1)\,
\label{shockK}
\eeq 
confirming their $\delta$-function behavior required by Penrose
\cite{Penr2}. The Sturm-Liouville eqn~(\ref{KSL}) is  satisfied when an appropriate regularization is chosen \cite{Steinbauer,KunStein}.
The amplitude of the wave is recovered as
\beq
k=\frac{1}{2}\int_{-\infty}^{+\infty}\!\!\cA(U)dU.
\label{NormA}
\eeq
Henceforth we will choose $k=1/2$ for convenience in all figures.

Considering the $\delta$-function profile (\ref{shockK}) has far-reaching consequences. Let us integrate the Sturm-Liouville eqn $\ddot{P}=KP$ in (\ref{KSL}) over an interval $U_i < 0 < U_f$ with $U_i$ and $U_f$ chosen arbitrarily in the before and in the after-zones, respectively,
\beq
\dot P(U_f)-\dot P(U_i)=
\int_{U_i}^{U_f} K(u)P(u)du
=
c_0
\label{genPjump}
\eeq
by (\ref{genBrink}) and (\ref{shockK}) and the fact that $P(0)=\bone$. 
Then letting $u_i\to 0-$ and $u_f\to 0+$ allows us to conclude that the $\delta$-function at the origin makes  \emph{$\dot{P}$ jump} at $u=U=0$,
\beq
\Delta \dot{P}\equiv
\dot P(0+)-\dot P(0-)=
\dot P(0+)=c_0.
\label{Pjump}
\eeq

Jumps are characteristic  of impulsive waves:  if $\cA$ was  smooth and thus bounded then there would be no jump. Such jumps will play a crucial r\^ole in what follows.

%%%%%%%%%%%%%%%%%%%%%%%%%%
\subsection{Souriau (S) coordinates}\label{Ssection}
%%%%%%%%%%%%%%%%%%%%%%%%%%

Besides  the widely used B  and  BJR  coordinates, 
 $(\bX, U,V)$ and $(\bx, u,v)$,  respectively, one also has yet another  coordinate system \cite{Sou73} whose use is particularly convenient in the flat case.
Start with a sandwich Brinkmann pp-wave written in BJR coordinates as in (\ref{BJRmetrics}).
The most general form of metrics in the flat zones is given by
\cite{Sou73,LongMemory}
\begin{equation}
%\medbox{
a(u)=a_0^\half\left(\bone+(u-u_0)c_0\right)^2a_0^\half
%}
\label{a}
\end{equation}
with $c_0=\half{}a_0^{-\half}\da_0\,a_0^{-\half}$, where $a_0=a(u_0)$ and $\da_0=\dot{a}(u_0)$ are initial conditions for some value~$u_0$
chosen within the flat region;  here $a_0^\half$ is a square-root of the matrix $a_0$.
Then the change of coordinates $(\bx, u,v)\to(\hbx,\hu,\hv)$ given by
\begin{subequations}
\begin{align}
\label{hxx}
\hbx
&=
\left(\bone+(u-u_0)c_0\right)a_0^\half\bx,
\\
\hu &=u,
\\
\label{hvv}
\hv&=
v-\half\bx\cdot{}a_0^\half{}c_0\left(\bone+(u-u_0)c_0\right)a_0^\half{}\bx,
\end{align}
\label{Scoords}
\end{subequations}
brings  the metric (\ref{BJRmetrics}) in the flat zone of spacetime we consider to  the Minkowski form, namely
\begin{eqnarray}
\rg=d\bx\cdot{}a(u)d\bx+2du\,dv
=d\hbx\cdot{}d\hbx+2d\hu\,d\hv.
\label{Minkowski}
\end{eqnarray} 
The ``hatted'' coordinates $(\hbx,\hu,\hv)$ in terms of which the metric is manifestly flat will be referred to as \emph{Souriau coordinates} \cite{Sou73}. The inverse of the coordinate change (\ref{Scoords}) is  
\begin{subequations}
\begin{align}
\label{xhx}
\bx
&=a_0^{-\half}\left(\bone+(u-u_0)c_0\right)^{-1}\hbx,
\\
u&=\hu
\\
\label{vhx}
v&=\hv+\half\hbx\cdot{}c_0\left(\bone+(u-u_0)c_0\right)^{-1}\hbx.
\end{align}
\label{iScoords}
\end{subequations}
To comply with our assumptions for impulsive gravitational waves, we will put from now on $u_0=0$ and $a_0=\bone$ so that the transformation formul{\ae} (\ref{Scoords}) and  (\ref{iScoords}) become
\beq
\label{Phxx}
\barraynb{lll}
\hbx
&=&
P(u)\bx,\\
\hv&=&v-\half\bx\cdot c_0P(u)\bx
\earraynb
\qquad\Leftrightarrow\qquad
\barraynb{lll}
\bx&=&P^{-1}(u)\hbx,\\
v&=&\hv+\half 
\hbx\cdot c_0P^{-1}(u)\hbx,
\earraynb
\eeq
respectively, with $P$ given as in (\ref{Pbeforeafter}), completed with $\hu=u$ as before.
\goodbreak

%%%%%%%%%%%%%%%%%%%%%%%%%%%%%%%%%%%%%%%%
\section{Impulsive wave as limit of Gaussians}\label{Gaussians}
%%%%%%%%%%%%%%%%%%%%%%%%%%%%%%%%%%%%%%%%

The form  (\ref{shockK}) suggests that the impulsive-wave profile could be obtained, in Brinkmann coordinates, by squeezing a smooth Gaussian profile  to a Dirac $\delta$. Let us indeed consider
\beq
\cA_\lambda(U)=\frac{
 \lambda}{\sqrt{\pi}}\,e^{-\lambda^2U^2}
\label{lambdacA}
\eeq
normalized as $\int_{-\infty}^{+\infty}\cA_\lambda(U)dU=1$ (consistently with our choice $k=\half$ made for figures).
Then we can calculate  the corresponding matrices $P_\lambda$  and $a_\lambda(u)$ numerically. FIGs.\ref{adeltalimit} and \ref{Pdeltalimit}
 confirm that, when  squeezing the Gaussians to a Dirac $\delta$-function by letting 
$\lambda\to\infty$,  the components of $P_{\lambda}(u)$ and of the transverse metric 
$a_\lambda(u)=P^T_{\lambda}(u)P_{\lambda}(u)$ tend to those of the impulsive wave, namely (\ref{abeforeafter}) and (\ref{Pbeforeafter}).
 
For finite $\lambda$ the coordinate transformation (\ref{BBJRtrans})
between B and BJR coordinates is smooth. However in the limit $\lambda\to\infty$, its $\bX$-part, while still  continuous, becomes non-differentiable as shown in  FIGs.\ref{adeltalimit} and \ref{Pdeltalimit}. For $V$, it is not even continuous  at $U=0$, 
\beq
V_-=v_0 
\qquad\text{and}\qquad
 V_+=
v_0-\frac{1}{2}\bX_0\cdot{c_0}\,\bX_0.
\label{Vjump}
\eeq\,
The additional term here corresponds to the ``gluing'' of  Penrose, eqn (2)  of \cite{Penr2};  cf. also eqn (3) of \cite{Steinbauer}.

%%%%%%%%%%%%%%%%%%%%%%%%%%%%%%%%%%%%%%%%%%%%%%%%%%%%%%
\section{Interlude: Carroll symmetry}\label{Carrollsec}
%%%%%%%%%%%%%%%%%%%%%%%%%%%%%%%%%%%%%%%%%%%%%%%%%%%%%%

%%%%%%%%%%%%%%%%%%
\subsection{Isometries in Baldwin-Jeffery-Rosen coordinates}
%%%%%%%%%%%%%%%%%%

The isometry group of a \emph{smooth generic} plane gravitational wave has long been known to be a $5$-dimensional Lie group \cite{Sou73,BoPiRo,Gibb75,BoPi89,Carroll4GW,Torre}. For \emph{any} transverse matrix $a(u)$, its action on spacetime is explicitly described in BJR coordinates  \cite{Sou73,Carroll4GW} by  
\begin{subequations}
\begin{align}
\label{Carrollx}
\bx&\to\bx+H(u)\,\bb+\bc,
\\
u&\to u,
\\
v&\to v-\bb\cdot\bx - \2\bb\cdot{}H(u)\,\bb+f,
\label{Carrollv}
\end{align}
\label{genCarr}
\end{subequations}
with $\bb,\bc\in\bbR^2$ and $f\in\bbR$,  where 
\beq
H(u)=\int_0^u{\!\!a^{-1}(w)dw}
\label{Hmatrix}
\eeq
is a primitive of $a^{-1}(u)$ which we choose to vanish at $u=0$, say.
The isometries of the metric (\ref{BJRmetrics}) form therefore a group isomorphic to the group of matrices
\beq
A=\barray{ccc}
\bone&0&\bc
\\
-\bb^T&1&f\\
0&0&1
\earray,
\label{Isom}
\eeq
identified with
the \emph{Carroll group in $2+1$ dimensions ``without rotations''} \cite{Carroll4GW}. This group is actually isomorphic to the commutator subgroup $\left[\mathrm{Carr}(2+1),\mathrm{Carr}(2+1)\right]$ of the full Carroll group \cite{JMLL} in $2+1$ dimensions.
\goodbreak

In the impulsive case, $H(u)$ be calculated analytically \emph{separately both in the before and in the after-zone}
 starting from the same point $u_0=0$.
 Using (\ref{abeforeafter}) we find $H(u)=u\,\bone$ for $u\leq0$, so that
boosts act conventionally in the before-zone, $\bx \to \bx+ u\,\bb$
for $u\leq0$.
However for $u > 0$ we have 
$H(u)=c_0^{-1}\left(\bone-(\bone+u\,c_0)^{-1}\right)$, 
yielding the general expression 
%$H(u)=c_0^{-1}\left(\bone-(\bone+u\,\theta(u)\,c_0)^{-1}\right) + u\left(1-\theta(u)\right)\bone$, i.e.,
\beq
\medbox{
H(u)=c_0^{-1}\left(\bone-(\bone+u\,\theta(u)\,c_0)^{-1}\right) + u\left(1-\theta(u)\right)\bone=u_+\,P^{-1}(u)+u_-\,\bone
}
\label{Hbeforeafter}
\eeq
with $P(u)$ as in (\ref{Pbeforeafter}), using the shorthand
$%\beq
u_+=u\,\theta(u), \;
u_-=u\left(1-\theta(u)\right).
%\label{upm}
$ %\eeq
Therefore boosts act non-conventionally in the after-zone, namely as
\beq
\bx \to
\bx+u\, P^{-1}(u)\,\bb\,.
\label{afterboost}
\eeq
Similarly, translations $\bx \to \bx+\bc$, which can be seen as natural translations in BJR coordinates, are symmetries in any flat zone. 

The Carroll action (\ref{genCarr}) involves the \emph{integral} (\ref{Hmatrix});  this action is defined and is continuous over the entire space~: the integration smoothes out the breakings of the metric.

Henceforth we restrict ourselves to the Lie algebra of the isometry group.

It is worth mentioning that the ``distorted'' action (\ref{genCarr}) of the Carroll group can also be derived using S coordinates, in terms of which the metric takes a Minkowski form. 
The $(2+1)$-dimensional Carroll group is indeed a subgroup of the Poincar\'e group in $3+1$ dimensions \cite{Bargmann,Carroll4GW};  its action on S coordinates is the restriction to Carroll of the usual Poincar\'e action on Minkowski space 
\cite{Sou73}.  Boosts, for example,  act  conventionally on the ``hatted'' coordinates,
\beq
\hbx \to \hbx + u\,\bb.
\eeq
Expressing this action in terms BJR coordinates using (\ref{hxx}) and  (\ref{Hbeforeafter}) we recover (\ref{genCarr}).

%%%%%%%%%%%%%%%%%%
\subsection{The Lie algebra of infinitesimal isometries in Brinkmann coordinates}
%%%%%%%%%%%%%%%%%%

The infinitesimal symmetries of a plane gravitational wave can  also be determined in Brinkmann coordinates $(X^1,X^2,U,V)$. The  Killing vectors are indeed of the form
\begin{equation}
Z=P_1(U)\,\partial_1+P_2(U)\,\partial_2+h(\bX,U)\,\partial_V,
\label{Z}
\end{equation}\vskip-3mm
where\vskip-10mm
\begin{subequations}
\begin{align}
\label{fppgpp}
\ddot{P}_1(U)&=+\half\cA(U)P_1(U),
\qquad
\ddot{P_2}(U)=-\half\cA(U)P_2(U),
\\
\label{h}
\;\;
h(X^1,X^2,U)&=-X^1\dot{P_1}(U)-X^2\dot{P}_2(U)+\eta
\;
\end{align}
\label{BKilling}
\end{subequations}
with $\eta=\const$ The two Sturm-Liouville equations (\ref{fppgpp})  here yield a $4$-parameter family of solutions. For our $\delta$-function choice  (\ref{shockK}) the integration of these equations is elementary, and yields the general solutions 
\beq
P_1(U)=\left[1+k\,U\theta(U)\right]\beta_1+U\gamma_1,
\qquad
P_2(U)=\left[1-k\,U\theta(U)\right]\beta_2+U\gamma_2
\label{fgU}
\eeq
with $\beta_1,\beta_2,\gamma_1,\gamma_2$ integration constants. The last component, $h(\bX,U)$, is then determined by (\ref{h}), 
namely
\beq
h(\bX,U)=-\bX\cdot\left(\theta(U)\,c_0\,\bbeta+\bgamma\right)+\eta.
\eeq
The infinitesimal isometries of our metric (\ref{Bplanewave}) form thus a $5$-dimensional Lie algebra, which is clearly isomorphic to the Lie algebra of the matrix group (\ref{Isom}). Remarkably,
eqn (\ref{fgU}) can also be written as
\beq
\barray{c}
P_1 \\ P_2
\earray
= P(U)\bbeta+U\bgamma
\eeq
where, as anticipated by our notations, $P_1=P_{11}$ and $P_2=P_{22}$ are precisely the components of the diagonal matrix $P$ in (\ref{Pbeforeafter}). In fact, this is not an accident~: the coefficients of the Killing vectors are, for \emph{any} plane gravitational wave, solutions of eqn (\ref{KSL})
 \cite{BoPiRo,Torre}. What distinguishes our case here from the general one is that, in the impulsive case, eqn (\ref{KSL}) can be solved explicitly as (\ref{Pbeforeafter}), whereas no explicit solution is known for a generic Sturm-Liouville equation. This may well be the reason why the algebraic structure
of the symmetry algebra has been identified only recently as that of the \emph{Carroll Lie algebra with no rotations} in $2+1$ dimensions, 
with $\bbeta$ and $\bgamma$ generating boosts and space translations respectively, and $\eta$ ``vertical time'' translations \cite{Sou73,Carroll4GW}, cf. footnote 2.

%%%%%%%%%%%%%%%%%%%%%%%%%%%%%%%%%%%%
\subsection{Geodesics: the Noether and Jacobi constants of the motion}
%%%%%%%%%%%%%%%%%%%%%%%%%%%%%%%%%%%%

As preparation to the study of free fall in an impulsive plane gravitational wave, let us recall the form of the constants of the motion associated with the symmetries of the problem. These first-integrals will prove crucial to integrate in elementary terms the equations of geodesics of the metric (\ref{BJRmetrics}) in its \emph{impulsive} guise (\ref{abeforeafter}).

\goodbreak

The isometry group (\ref{Isom}) generates $5$ conserved quantities by Noether's theorem applied to geodesic motion. For geodesics $(\bx(s),u(s),v(s))$,
they are, \emph{in the before- and the after-zones} \emph{separately}, given  by \cite{Sou73,Carroll4GW}
\beq
\bp_{\pm} = a(u)\,\frac{d\bx_{\pm}}{\!\!ds},
\qquad
\bk_{\pm}=\bx_{\pm}(u)\frac{du}{ds}-H(u)\,\bp_{\pm},
\qquad
\mu_\pm=\frac{du}{ds},
\label{CarCons}
\eeq
where the $\pm$ refers to $u\leq0$ and $u>0$, respectively. 
For causal \emph{continuous} geodesics, where
\beq
e_\pm=\half\rg_{\mu\nu}\frac{dx_\pm^\mu}{\!\!ds}\frac{dx_\pm^\nu}{\!\!ds}=\const\leq0
\label{epm}
\eeq
we may put $\mu_\pm=1$ and hence use from now on $s=u$ as curve parameter.

Since the associated conserved quantities (\ref{CarCons}) involve the motion and in particular the velocity, they may jump. The Noether quantities $\bp_\pm$ and $\bk_\pm$ and the ``Jacobi'' constants of the motion, $e_\pm$, may  indeed be different before and after the impulse, as will be shown below.

%%%%%%%%%%%%%%%%%%%%%%%%%%%%%%%%%%%%%%%%%%%%%%%%%%%%%%
\section{Impulsive waves: geodesics}\label{GeoSec}
%%%%%%%%%%%%%%%%%%%%%%%%%%%%%%%%%%%%%%%%%%%%%%%%%%%%%%

Now we turn to the geodesics which have a  subtle behavior, due, precisely, to the jumps, characteristic  of the $\delta$-function profile.

%\goodbreak
%%%%%%%%%%%%%%%%%%%%%%%%%%%%%%%
\subsection{Geodesics in Baldwin-Jeffery-Rosen and in Souriau coordinates }\label{BJRGeosec}
%%%%%%%%%%%%%%%%%%%%%%%%%%%%%%%

Test particle trajectories identified with the geodesics of the plane GW metric can be determined analytically in both of the flat before and after-zones $u\le0$ and $u>0$ {separately}, distinguished by $\mp$ indices. A simple way to find them is to use the conservation laws written in BJR coordinates   \cite{Sou73,Carroll4GW,ShortMemory,LongMemory}.
From the expression of $\bp_\pm$ in (\ref{CarCons}) and from eqn (\ref{epm}) we infer that 
\beq
e_\pm=\half\bp_\pm\cdot{}a^{-1}(u)\,\bp_\pm+\dot{v}_\pm(u),
\eeq
 hence (\ref{CarCons}) leaves us with 
\begin{subequations}
\begin{align}
\bx_{\pm}(u)&=\bk_\pm+H(u)\,\bp_{\pm},
\label{xpm1}
\\
v_{\pm}(u)&=-\half \bp_{\pm}\cdot H(u)\,\bp_{\pm} + {\rm e}_{\pm}\,u+ d_{\pm},
\label{vpm1}
\end{align}
\label{BJRtraj}
\end{subequations}
where we anticipated that the quantities $\bp_{\pm},\bk_\pm,d_\pm$, and $e_{\pm}$, conserved in their respective zones, may (unlike in the smooth case), be \emph{different}. 

\goodbreak

Now, these geodesics are meant to represent worldlines of particles in each zone: they must be continuous functions of $u$, implying that 
\beq
\bk_\pm=\bx(0)=\bx_0
\qquad\text{and}\qquad d_\pm=v_\pm(0)=v_0
\eeq 
since~$H(0)=0$.
Moreover, we have 
\beq
\bp_\pm=\dot\bx_\pm(0)
\qquad\text{and}\qquad 
\dot{v}_\pm(0)=e_\pm-\half\vert\bp_\pm\vert^2.
\eeq
 
At this stage, the latter constants of integration remain arbitrary as they parametrize independent geodesics in each half-zone. Thus
we end up with the parametrized continuous geodesics
\begin{subequations}
\begin{align}
\bx_{\pm}(u)&=H(u)\,\dot\bx_\pm(0)+\bx_0, 
\label{xpm}
\\
v_{\pm}(u)&=\half\dot\bx_{\pm}(0)\cdot\left[u-H(u)\right]\dot\bx_{\pm}(0) + u\,\dot{v}_{\pm}(0)+ v_0,
\label{vpm}
\end{align}
\label{BJRgeodesics+-}
\end{subequations}
where $H(u)$ is as in (\ref{Hbeforeafter}). We note that the Jacobi constants read 
\beq
e_\pm=+\half\vert\dot\bx_\pm(0)\vert^2+\dot{v}_\pm(0).
\eeq
Using our previous notations $u_\pm$, we can write alternatively
\begin{equation}
\left\{
\begin{array}{ll}
\bx(u)&=u_+\,P^{-1}(u)\,\dot\bx_+(0)+u_-\,\dot\bx_-(0)+\bx_0,
\\[5pt]
\displaystyle
v(u)&=\half\dot\bx_+(0)\cdot{}u_+\left(\bone-P^{-1}(u)\right)\dot\bx_+(0)+u_+\,\dot{v}_+(0)+u_-\,\dot{v}_-(0)+v_0,
\end{array}\right.
\label{BJRgeodesics}
\end{equation}

%%%%%%%%%%%%%%%%%%%%%%%%%%%%%%%
%\subsection{Free fall of particles initially at rest (S and BJR coordinates)}\label{BJRfreeFallSection}
%%%%%%%%%%%%%%%%%%%%%%%%%%%%%%%
Henceforth, we limit our investigations at particles initially at rest in the before zone.

Amongst the previous solutions, what are the physical ones suited to the description of free fall in a plane impulsive GW? Indeed, the worldlines of particles in $3+1$ dimension should be characterized by $4$ initial positions $\bx_0,v_0$, and velocities $\dot\bx_0,\dot{v}_0$. So, how could we eliminate one of the spurious velocities, $\dot\bx_+(0)$, say? 
An answer is obtained by using the $S$-coordinates of sec. \ref{Ssection}
proposed by Souriau \cite{Sou73}.

As explained before, the metric outside the wave zone can be cast into a canonical Minkowskian form (\ref{Minkowski}) in either of the flat zones.
The coordinate transformation (\ref{Phxx}) between the S and BJR coordinates in the after-zone $u>0$ of our impulsive wave  is,
\begin{equation}
\bx=\left(\bone+u\,c_0\right)^{-1}\hbx,
\qquad
u=\hu,
\qquad
v=\hv+\half\hbx\cdot{}c_0\left(\bone+u\,c_0\right)^{-1}\hbx
\label{S2BJR}
\end{equation}
 where $c_0\neq0$.
The metric $\rg=d\hbx\cdot{}d\hbx+2d\hu\,d\hv$  no longer involves $c_0$ [which got hidden in the transformation]. Formally  the same transformation holds therefore in the before-zone characterized by $c_0=0$, where it is in fact  is the identity, 
$ \bx=\hbx,
\,
u=\hu,
\,
v=\hv
$ 
wherever $u\leq0$.

\goodbreak

Consider now \textit{particles initially at rest} whose geodesics are clearly given in S coordinates by \textit{affine} parametric equations in the before-zone, namely
\begin{equation}
\hbx(\hu)=\hbx_0,
\qquad
\hv(\hu)=\hv_0+e\,\hu,
\label{GeodesicsBeforeZone}
\end{equation}
where $e=\const$ is the Jacobi first integral of the geodesic equations and $\hbx_0$, $\hat{v}_0$ constants of integration.
Now, the after zone being also flat and indeed Minkowskian when the hatted S coordinates are used,
%and the worldlines  \textit{continuous},
  we argue that the latter have the \emph{same} parametric form in the after-zone: (\ref{GeodesicsBeforeZone}) holds for all $u$.
Translating  (\ref{GeodesicsBeforeZone}) to the {original} BJR coordinates 
we find, using
$ 
\hbx_0=\bx_0,
\;
\hv_0=v_0-\half\bx_0\cdot{}c_0\bx_0
$ 
cf. (\ref{S2BJR}), 
the explicit parametric expression of the geodesics  in BJR coordinates \footnote{Equation (\ref{FreeFallinGW})  holds also for \textit{null geodesics}, $e=0$}
\begin{equation}
\medbox{
\left\{
\begin{array}{lll}
\bx(u)&=&\left(\bone+u\,\theta(u)c_0\right)^{-1}\bx_0\\
&=&P^{-1}(u)\bx_0,
\\[6pt]
v(u)&=&-\half\bx_0\cdot\left(\bone-(\bone+u\,\theta(u)c_0)^{-1}\right)c_0\bx_0+e\,u+v_0
\\
&=&-\half\bx_0\cdot\left(\bone-P^{-1}(u)\right)c_0\bx_0+e\,u+v_0.
\end{array}
\right.}
\label{FreeFallinGW}
\end{equation}
for all $u\in(-\infty,k^{-1})$. 
Eqn (\ref{FreeFallinGW}) allows us to interpret the ``hatted'' S coordinates as the points of the trajectories at $u=0$ \footnote{This also hints at that correspondence between the BJR and S coordinates fails to be one-to-one at points where the trajectories meet --- i.e., at caustic points.} (completed with $u$ itself).

We see that the geodesic equation (\ref{FreeFallinGW}) is a special case of (\ref{BJRgeodesics}) where the after-zone initial velocity has been  fixed by the initial conditions $\bx(0)=\bx_0$ and $\dot\bx(0-)=0$, namely
\beq
\medbox{
\dot\bx(0+)=-c_0\bx_0.}
\label{dotx0+}
\eeq
The impulsive GW induces a [sort of]  \textit{``percussion''}
\cite{Sou73}, since
\begin{subequations}
\begin{align}
\label{Deltadotbx}
\Delta\dot\bx&=\dot\bx(0+)-\dot\bx(0-)=-c_0\bx_0,
\\
\label{Deltadotv}
\Delta\dot{v}&=\dot{v}(0+)-\dot{v}(0-)=-\half\vert{}c_0\bx_0\vert^2.
\end{align}
\label{percussion}
\end{subequations}
We contend that this canonically determined solution of the geodesic equation is germane to a deterministic description of the scattering of particles initially at rest by a plane impulsive GW. Equation (\ref{percussion}) provides us with a special instance of the \emph{Velocity Memory Effect}, which also includes the ``longitudinal'' velocity, $\dot{v}$.

Let us note for further record that all BJR trajectories are continuous~: this follows from  (\ref{FreeFallinGW}). In particular, the longitudinal coordinate $v(u)$ suffers \emph{no discontinuity}. 
 
%%%%%%%%%%%%%%%%%%%%%%%%%%%%%%%%%%%%% 
\subsection{Geodesics in Brinkmann coordinates}\label{BGeosec}
%%%%%%%%%%%%%%%%%%%%%%%%%%%%%%%%%%%%%

Now we describe our geodesics, again, \emph{independently} of sec \ref{BJRGeosec}, in Brinkmann coordinates. The geodesic equations are
\begin{subequations}
\begin{align}
&\ddot{X}^1 - \half{\cA}(U) X^1 = 0,
\label{geoX1}
\\[5pt]
&\ddot{X}^2 + \half{\cA}(U) X^2 = 0,
\label{geoX2}
\\[5pt]
&\ddot{V}+\frac{1}{4}\dot\cA(U)\big((X^1)^2-(X^2)^2\big) 
+ 
\cA(U)\big(X^1\dot{X}^1-X^2\dot{X}^2\big)=0.
\label{geoV}
\end{align}
\label{Bgeoeqn}
\end{subequations}
In ``Bargmann terms" \cite{Bargmann,LongMemory}   (\ref{geoX1}) and (\ref{geoX2}) describe a time-dependent anisotropic ``oscillator'' in transverse space which, (assuming $\cA(U)>0$), is attractive in the $X^2$ coordinate and repulsive in the $X^1$ coordinate. 

Now, repeating the argument given in sec. \ref{planewavesec} for $\dot{P}$,
we show  that the transverse \emph{velocity}, $\dot{\bX}$, necessarily \emph{jumps} at $U=0$. To see this, we assume that the particle is at rest in the before-zone, 
$\dot{\bX}(U)=0$ for $U<0$, and
integrate (\ref{geoX1}) and (\ref{geoX2}) for the impulsive profile
${\cA}= 2k\,\delta(U)$ cf. (\ref{shockK}) over an interval 
$U_i < 0 < U_f$ which contains the origin. 

Assuming that standard distributional identities hold we get 
$\bX(U)=k\,U\theta(U)\bA+U\,\bB+\bC$ for some constants $\bA,\bB,\bC$. Plugging this into (\ref{geoX1}) and (\ref{geoX2}) yields $\bC=\bA$. Thus $\dot{\bX}=k\,\theta(U)\bA+\bB$, hence $\bB=0$ in view of our assumption 
$\dot\bX(0-)=0$. At last, we find $\bA=\bX(0-)=\bX_0$.
It follows from 
\beq
\dot\bX(U)=\theta(U)c_0\bX_0
\label{BdotX}
\eeq 
that the initial velocity of the Brinkmann trajectory jumps, \footnote{Note that $\bX$ and $P$ satisfy identical equations. \emph{The sign of the $\bX$-jump is the opposite} of that in BJR coordinates  in (\ref{Deltadotbx}).}
\beq
\medbox{
\dot{\bX}(0+)
=+c_0\,\bX_0\,.
}
\label{BXveljump} 
\eeq
The general  form of the spatial trajectory 
in a flat region is therefore \footnote{
The integration of eqn (\ref{geoV}) for $V(U)$ --- containing multiplication of distributions --- would require more elaborate techniques, 
see, e.g., \cite{Steinbauer}, which go beyond our scope here. 
%Curiously, integration would be possible if the coefficient of the second term was $1/2$ instead of $1/4$.
} 
\beq
\medbox{
\bX(U) = \left(\bone+U\theta(U)\,c_0\right)\bX_0=P(U)\bX_0.
}
\label{BX}
\eeq

\goodbreak

\begin{figure} [h]
\includegraphics[width=0.58\textwidth]{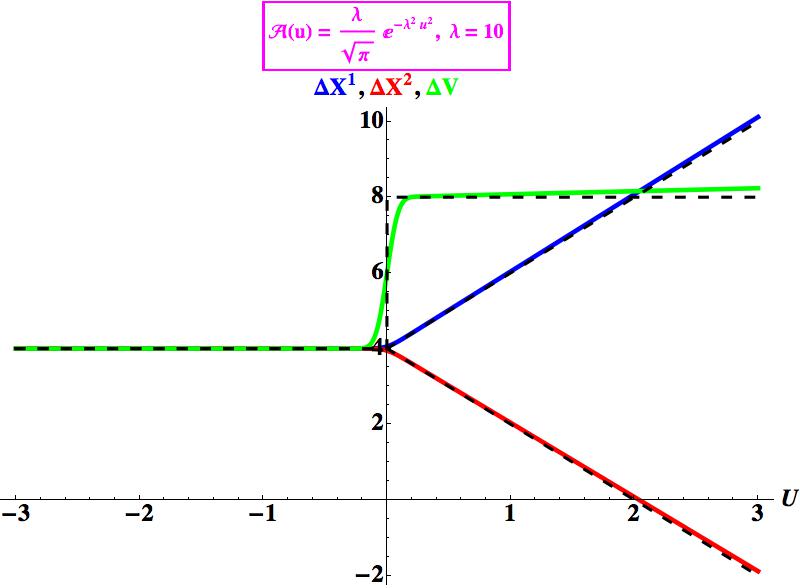}
 \vskip-5mm
\caption{\textit{The geodesics for Gaussian profile $\cA_{\lambda}$  exhibit, as does their $\lambda\to\infty$ limit,
 the \underline{Velocity Memory Effect}: after the wave has passed, the relative transverse position $(\blue{\Delta{X}^1}, \red{\Delta{X}^2})$ of particles initially at rest diverge apart, consistently with (\ref{Breldistvel}). The relative coordinate along the propagation, $\dgreen{\Delta{V}}$), changes sharply and letting $\lambda\to\infty$ it becomes discontinuous: the trajectory suffers a forward jump given by the difference of the respective expressions in (\ref{Vdiff}). Our plot corresponds to the initial conditions
$\blue{X^1_ 0 = 1}, \red{X^2_0 = 3}, \dgreen{V_0 = 2}$, 
$\blue{Y^1_ 0 = -3}, \red{Y^2_0 = -1}, \dgreen{W_0 = -2}$. 
}}
\label{B3geo}
\end{figure}

The transverse velocity of a particle at rest in the before-zone is, in the afterzone, $U$-dependent. More precisely, it  depends linearly on the initial position ${\bX}_0$,\footnote{Eqn (\ref{BdotX}) is consistent with eqn (II.19) of \cite{ShortMemory}.} as shown by the {black dashed} lines in Fig.\ref{B3geo}.
Consistently with (\ref{BXveljump}) and (\ref{BX}), the trajectories first focus in the attractive coordinate \red{$X^2$}; then, after passing the caustic point  at  $u_1=k^{-1}$, they diverge with slopes proportional to their initial positions. The repulsive coordinates \blue{$X^1$} diverge from the beginning. 
The \emph{relative distance between two trajectories grows therefore linearly with constant but non-zero relative velocity},
\begin{subequations}
\begin{align}
|\bX(U)-\bY(U)|&=\left|(\bX_0-\bY_0)+U\,c_0(\bX_0-\bY_0)\right|
\label{Breldist}
\\
|\dot{\bX}(U)-\dot{\bY}(U)|&= 
\left|c_0\big({\bX}_0-{\bY}_0\big)\right|=\const\neq0\,
\label{Brelvel}
\end{align}
\label{Breldistvel}
\end{subequations}
since for $k\neq0$ the  matrix $c_0$ is invertible.
We conclude that impulsive wave behave as
their smooth counterparts do \cite{ShortMemory,LongMemory}~:
 \emph{no permanent transverse}  \emph{displacement is possible};  
 they exhibit instead  the \emph{Velocity Memory Effect} \cite{BraGri,BraTho,GriPol,BoPi89,ShortMemory,LongMemory}. 
%%%%%%%%%%%%%%%%%
%%%%%%%%%%%%%%%%%

An even more dramatic effect is a \emph{discontinuity 
suffered by the ``vertical'' coordinate, $V$}.
 The Jacobi constant is now 
$
e = \half |\dot\bX|^2 + \dot{V} + \half \delta(U) \bX(U)\cdot c_0 \bX(U).
$ 
Integrating this expression between $U_i<0$ and $U_f>0$, and using
(\ref{BX}) and (\ref{BdotX}), we find
\begin{equation}
e[U_f - U_i] 
=
\half U_f\,\bX_0\cdot c_0 \bX_0
+V(U_f) - V(U_i)
+\half\bX_0\cdot c_0\bX_0.
\end{equation}
 In the limit $U_i\to 0-$ and $U_f\to 0+$ we end thus up with
\beq
V(0+) - V(0-) = -\half \bX_0\cdot c_0\bX_0.
\label{Vdiff}
\eeq 

%%%%%%%%%%%%%%%%%%%%%%%%%%%%%%%%%%%%%%%%%%%%%%%%%%
\subsection{Comparison of the trajectories in Brinkmann and in Baldwin-Jeffery-Rosen coordinates}
%%%%%%%%%%%%%%%%%%%%%%%%%%%%%%%%%%%%%%%%%%%%%%%%%

We conclude this section by relating the \emph{trajectories}  in  B and in BJR coordinates.
The naive expectation might be that this could be achieved by using the  transformation formula between the coordinates, (\ref{BBJRtrans}), i.e.,
\beq
\bX(U)=P(U)\,\bx(u),
\label{naiveXx}
\eeq
which is indeed correct in the case of continuous wave profiles for particles initially at rest, \cite{ShortMemory,LongMemory}, for which $\bx(u)=\bx_0=\const$ for all $u$. However,
identifying the initial positions,
$ 
\bx_0=\bX_0
$ 
and  combining (\ref{BX}) and (\ref{FreeFallinGW})
yields instead, 
\beq
\medbox{
\bX(U)= (P^T\!P)(u)\, \bx(u) = a(u)\, \bx(u)\,.\,
}
\label{goodXx}
\eeq
Where does the extra $P$-factor come from ? The clue is that 
\emph{ the delta-function $\delta(u)$  makes the velocity jump}  both in B and BJR coordinates  --- and does it in the \emph{opposite} way, see in (\ref{BXveljump}) and (\ref{Deltadotv}), respectively. 
The extra $P$ factor takes precisely care of these jumps~:
 the first $P$ in (\ref{goodXx}) straightens the trajectory (\ref{FreeFallinGW}) to the trivial one, 
$
P(u)\bx(u)=\bx_0,
$
which has zero initial BJR velocity  as in the smooth case \cite{ShortMemory,LongMemory};
 then the second $P(u)$ factor curls it up according to (\ref{naiveXx}), yielding $\bX(u)$ in (\ref{BX}).
 
 Deriving  w.r.t. $u>0$ and using (\ref{Pjump}) and (\ref{Deltadotbx}) confirms also that (\ref{goodXx}) flips over  the initial velocities, 
\beq
\dot{\bx}(0+)=-c_0\bx_0  \Rarrow  \dot{\bX}(0+)= + c_0\bX_0,
\eeq
\goodbreak 

The formula (\ref{goodXx}) allows us to clarify yet another puzzle.
Naively, it would seem  that (\ref{FreeFallinGW}) would show \emph{no memory effect}, since the transverse-space distance  between two arbitrary trajectories $\bx(u)$ and $\by(u)$ in (\ref{FreeFallinGW}) is \emph{constant}, 
\beqa
{\Vert}\bx(u)-\by(u){\Vert}
&=&
\sqrt{(P^{-1}\bx_0-P^{-1}\by_0)\cdot (P^TP)(P^{-1}\bx_0-P^{-1}\by_0)}
=
{\vert}\bx_0-\by_0{\vert}=\const\,\qquad\quad
\label{naiveBJRdist}
\eeqa
The error comes from having forgotten that the \emph{true distance} is not (\ref{naiveBJRdist}) but the one between the corresponding Brinkmann trajectories,
\beqa
|\bX(U)-\bY(U)|&=&\sqrt{[P^2(u)\big(\bx(u)-\by(u)\big)]^2}=
\left\vert{}P(U)(\bx_0-\by_0)\right\vert
\label{trueBJRdist}
\eeqa
which grows affinely with $U$, cf. (\ref{Breldistvel}).

Turning to the conserved quantities, those
 associated with the symmetries in section \ref{Carrollsec} are obtained by the Noether theorem. Since the Killing vector in (\ref{Z}) has no component along $\p_{U}$, the problematic ``vertical velocity" $dV/dU$ drops out. Nor does any $\delta$-function show up, providing us with
\begin{subequations}
\begin{align}
\label{Bboostmom}
\bK&= \bX_0=P^{-1}(U)\bX(U)=P(u)\bx(u)=\bx_0=\bk,
\\
\bm{\Pi}_+&=\dot{\bX}(0+)=c_0\bX_0=-\dot{\bx}(0+)=\bp_+, 
\label{Pi+}
\\
\bm{\Pi}_-&=\dot{\bX}(0-)=0=-\dot{\bx}(0-)=\bp_-.
\label{Pi-}
\end{align}
\label{BConsQuant}
\end{subequations}
Our  formul{\ae} are consistent with the approach based on Souriau's $S$ coordinates, as seen by combining the B $\to$ BJR
and BJR $\to$ S maps in  (\ref{goodXx}) and  in (\ref{Pbeforeafter}), respectively,
\beq
\bX(U) = P^2(u)\,\bx=P(u)\,\hbx(u)
\Rarrow
\bx(u) = P^{-2}(u)\,\bX(u) =P^{-1}(u)\,\hbx(u)
\label{Xxhx}
\eeq
which confirms that the S coordinates are in fact the (common) initial positions at $u=0$.

%%%%%%%%%%%%%%%%%%%%%%%%%%%%%%%%%%%%%%
\section{Geodesics for Gaussian profile}\label{GaussGeosec}
%%%%%%%%%%%%%%%%%%%%%%%%%%%%%%%%%%%%%%

We have seen in sec.\ref{Gaussians} that the impulsive metric is obtained by shrinking Gaussians; now we turn to their geodesics. We emphasise that neither the before nor the after zone is rigorously defined in this case; we use the notation
$u<<0$ and $u>>0$ merely to indicate ``far-away regions, where the metric components are very small".

\begin{figure} [h]
\includegraphics[width=0.465\textwidth]{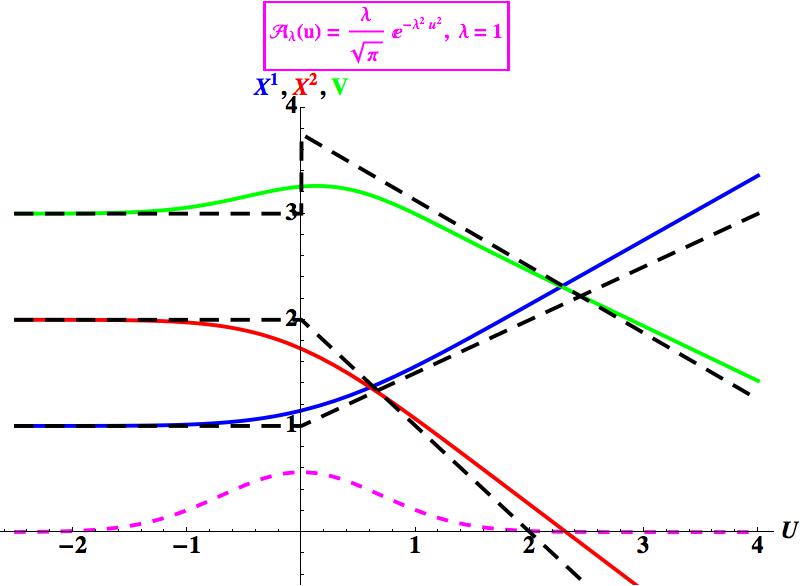}\;\;\;\;\;
 \includegraphics[width=0.465\textwidth]{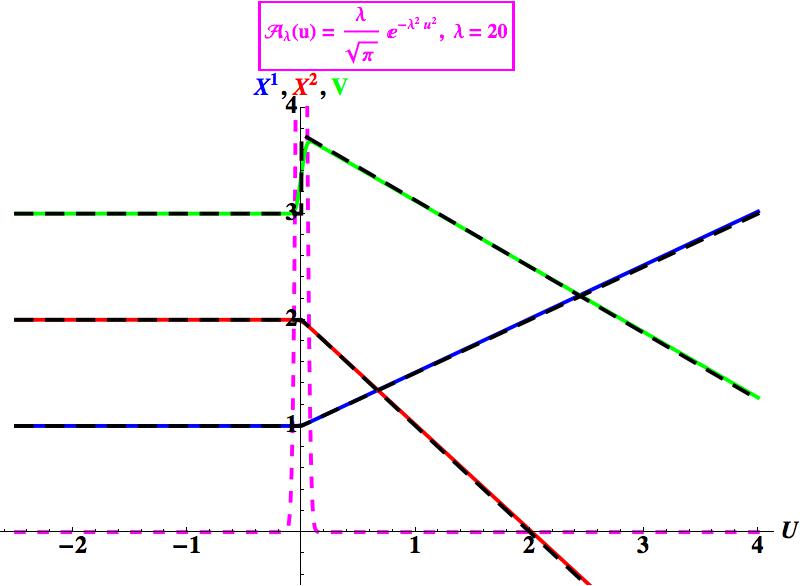}
\null\vskip-7mm
\null\hskip-20mm (a)\hskip75mm (b)
\vskip-5mm
\caption{\textit{In Brinkmann coordinates 
 the  geodesics $\big(\blue{X^1(U)}, \red{X^2(U)}, \dgreen{V}(U)\big)$  found numerically for the Gaussian profiles $\cA_{\lambda}$  
 tend, when $\lambda\to\infty$, to that in eqn (\ref{BX}), composed of broken dashed straight lines in black, valid  for the $\delta$-delta function. 
 In the impulsive case the transverse coordinates are $C^0$ but not $C^1$, whereas both $V$ and $\dot{V}$ jump. Our initial conditions are $\blue{X^1_0 = 1}$, $\red{X^2_0= 2}$ and $\dgreen{V_0 = 3}$
at $u= - \infty$. For all $\lambda$ there is a unique caustic, namely in the attractive sector $\red{X^2}$, close to the  impulsive value $u_1=2$.
}}
\label{squeezegeoB}
\end{figure}

%
%%%%%%%%%%%%%%%%%%%%%%%%%%%%%%%%%%%%% 
%\subsection{Geodesics in Brinkmann coordinates}\label{GGeoBsec}
%%%%%%%%%%%%%%%%%%%%%%%%%%%%%%%%%%%%%
%
 The equations (\ref{Bgeoeqn}) are valid for any profile including  Gaussians $\cA_{\lambda}=({\lambda}/{\sqrt{\pi}})\,\exp[{-\lambda^2U^2}]$ in (\ref{lambdacA}). However they can be solved only numerically; the results are depicted in Fig.\ref{squeezegeoB}.

%%%%%%%%%%%%%%%%%%
%%%%%%%%%%%%%%%%

\begin{figure} [h]
%\hskip-17mm
 \includegraphics[width=0.5\textwidth]{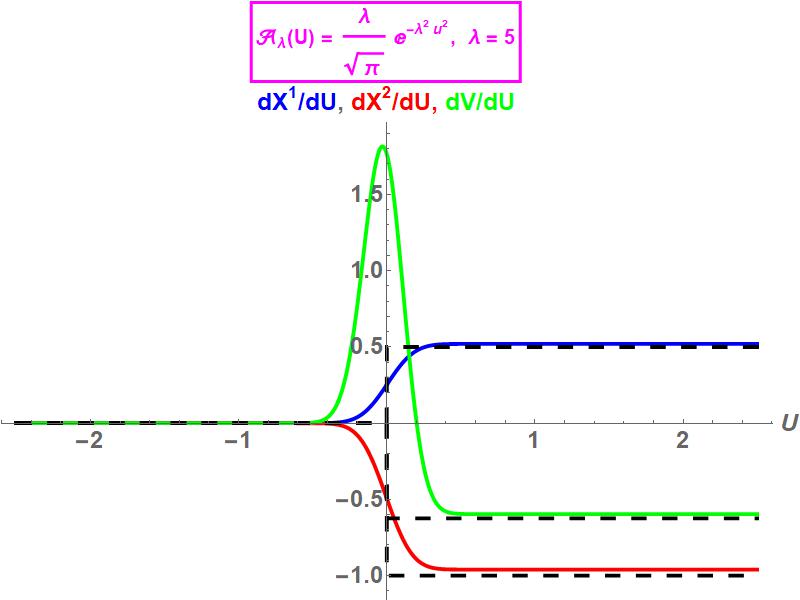}
 \vskip-5mm
\caption{\textit{The velocity calculated for a Gaussian tends to  non-zero constant value, consistent with the one in the impulsive limit $c_0\bX$ in (\ref{BXveljump}), (shown in  black dashed lines).
 The particles are initially at rest; their initial positions are $\blue{X^1_0=1}, \red{X^2_0=2}$.}}
\label{BvelXYV.jpg}
\end{figure}

For $U_i\ll0$ and $U_f\gg0$ (alternatively for large $\lambda$) our trajectories exhibit, once again, the \emph{Velocity Memory Effect}, as seen by integrating eqns (\ref{geoX1})-(\ref{geoX2}) over an interval
$[U_i,U_f]$, where these values are defined by the requirement that the Gaussian be very small outside the interval.
Then, since the components of $\bX$ satisfy the same equations as those of the diagonal matrix $P$, the proof of (\ref {genPjump}) yields the velocity jump
\beq
\Delta\dot{\bX}= \int_{U_i}^{U_f} \half\cA(U)\,\diag(1,-1)\bX(U) dU\,.
\label{Ddot}
\eeq
For a $\delta$-function profile this would be $c_0\bX_0$; for Gaussian profile it is somewhat different. How much~? It on depends on where the approximate sandwich values $U_i$ and $U_f$ are chosen and on how much $\bX(U)$ varies between them. Letting $\lambda\to\infty$,
(\ref{Ddot}) would converge to the $\delta$-function value $c_0\bX_0$.
For large $\lambda$ the  velocities tend rapidly to constant values,  as shown in Fig.\ref{BvelXYV.jpg}.\vskip-3mm
%%%%%%%%%%%%%%%%%%
%%%%%%%%%%%%%

A rigorous study of the behavior of $V(U)$ is more subtle, though~: the  procedure used for $\bX$ would require ill-defined multiplication of distributions, 
whose  handling \cite{Steinbauer,KunStein} goes beyond  the scope of this paper. Here we satisfy ourselves with our plots.

%%%%%%%%%%%%%%%%%%%%%%%%%%%%%%
%\subsection{Tissot diagrams}\label{Tissot}
%%%%%%%%%%%%%%%%%%%%%%%%%%%%%%

\begin{figure}[h]
\includegraphics[width=0.62\textwidth]{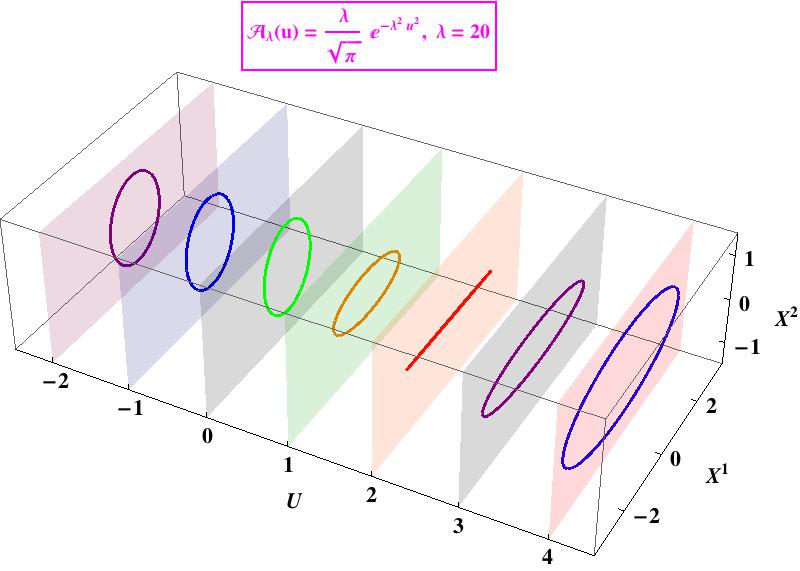}
\\
\vskip-4mm
\caption{\textit{Deformation of the initial Tissot circle for  squeezed Gaussian profile $\cA_{\lambda}=(\lambda/\pi)e^{-\lambda^2u^2}$ with $\lambda=20$. 
The only caustic arises for the attractive coordinate $X^2$ close to  the impulsive critical value $u_1=k^{-1}= 2$; the $X^1$ coordinates diverge apart all the time in the afterzone.
}}\vskip-4mm
\label{Tissotfig}
\end{figure}

The memory effect can nicely be illustrated using Tissot diagrams  borrowed from cartography \cite{MiThoWhe,ShortMemory,LongMemory}: one considers a tube of timelike geodesics starting from a circle of radius $R$ (say) in the transverse plane in the before-zone, see Fig.\ref{Tissotfig}.
%%%%%%%%%%%%%%%%%%
%%%%%%%%%%% 

So far, we only considered test particles initially at rest in the before-zone, $\dot{\bX}(U)=0$ for $u\leq0$. But this is by no means mandatory: our general solution works for any initial condition in any flat region \cite{Maluf}. Thus we should solve the geodesic equation separately in the before and in the after-zones with appropriate respective initial conditions and then glue them together at $U=0$ taking into account the jumping condition (\ref{BXveljump}), cf. Fig.\ref{BJRxv}.

\goodbreak
%%%%%%%%%%%%%%%%%%%%%%%%%%%%%%%%%%%%%%%%%
\section{Comparison with other approaches}\label{SteinbauerSec}
%%%%%%%%%%%%%%%%%%%%%%%%%%%%%%%%%%%%%%%%%

\begin{figure} [h]
 \includegraphics[width=0.43\textwidth]{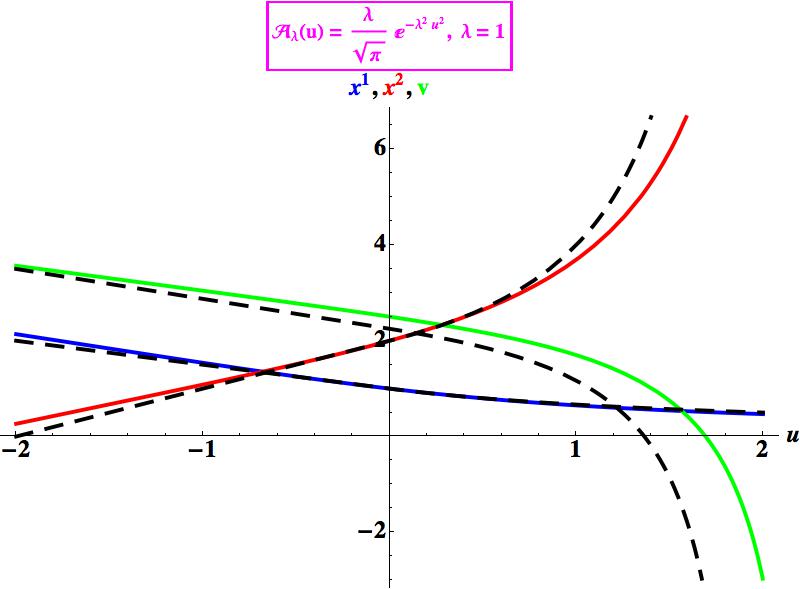} 
 \;\;\;\,\;±;
\includegraphics[width=0.43\textwidth]{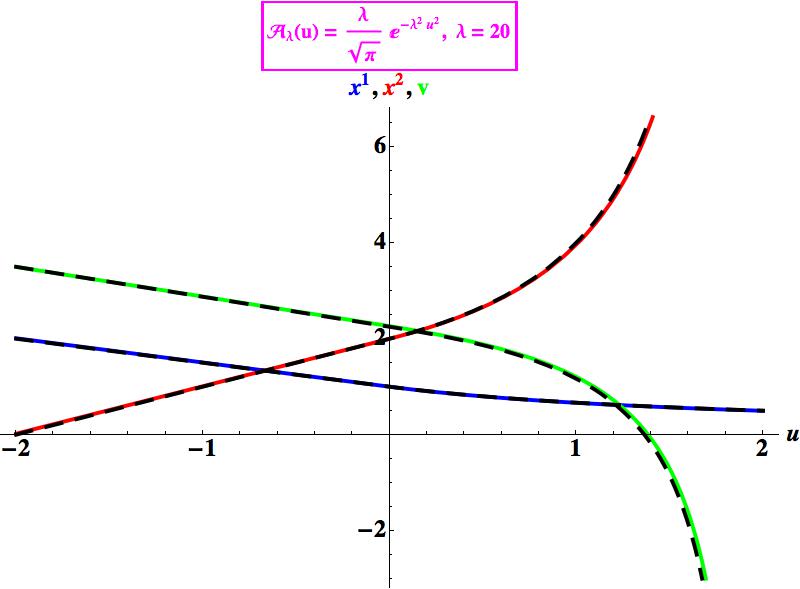}
\\
\hskip-2mm
(a)\hskip75mm (b)\\
 \vskip-3mm
\caption{\textit{The  BJR components of the geodesic  eqn (\ref{BJRPGeo})  solved numerically for the
 Gaussian profile $\cA_\lambda$. With appropriate initial conditions the trajectory tends, when $\lambda\to\infty$,  to our  (\ref{FreeFallinGW}) in the after-zone $u>0$ but \underline{not} in the before-zone; it tends instead globally to the Steinbauer solution (\ref{SB14}) for \emph{all $u$}. 
}}
\label{BJRxv}
\end{figure}

In \cite{Steinbauer} Steinbauer presented geodesics in the impulsive case. After correcting some typos, his equations \# (14) in \cite{Steinbauer} are
\beq
\left\{\begin{array}{lll}
x^1(u) &=& 
\dot{x}^1_0 \big(\displaystyle\frac{u_+}{1+u_+} + u_-\big)+x^1_0,
\\[14pt]
x^2(u) &=&
\dot{x}^2_0\,\big(\displaystyle\frac{u_+}{1-u_+} +  u_-\big)+x^2_0,
\\[14pt]
v(u) &=&
\half\,\left[(\dot{x}^1_0)^2 \displaystyle\frac{u_+}{1+u_+} - (\dot{x}^2_0)^2 \displaystyle\frac{u_+}{1-u_+}\right]+u\,\dot{v}_0 +v_0,
\end{array}\right.
\label{SB14}
\eeq 
where $u_\pm$ are as above. It is \emph{tacitly assumed that all curves are $C^1$} so that $\dot{\bx}_0$ is the common left-and-right velocity at $u=0$. It is shown in dashed black lines in Fig.\ref{BJRxv}.

Choosing suitable initial conditions, the Steinbauer solution (\ref{SB14}) reproduces \emph{either the first, or the second half, but not the entire broken trajectory  of our (\ref{BJRgeodesics})}.
Choosing $\dot{\bx}_0=0$ would yield indeed the trivial solution $\bx(u)=\bx_0$ which is fine in the before-zone, but not in the after-zone. (Note that particles at rest in the before-zone, 
$\dot\bx_0=0$, would remain at rest in the after-zone: no scattering would occur.) 
Putting instead $\dot{\bx}_0=-c_0\bx_0$ as in (\ref{dotx0+}) would yield our solution (\ref{BJRgeodesics}) in the after-zone, but \emph{not} in the before-zone. 

\goodbreak
For the sake of comparison, we present, with the help of (\ref{c00}), our parametrized $C^0$ geodesics in a coordinate-wise form similar to the Steinbauer expression (\ref{SB14}),
\begin{equation}
\left\{\begin{array}{lll}
x^1(u)&=&\,\dot{x}^1_+(0)\displaystyle\frac{u_+}{1+k\,u_+}+u_-\,\dot{x}^1_-(0)+x^1_0\,,
\\[14pt]
x^2(u)&=&\dot{x}^2_+(0)\displaystyle\frac{u_+}{1-k\,u_+}+u_-\,\dot{x}^2_-(0)+x^2_0\,,
\\[14pt]
v(u)&=&\half\,k\,u_+^2\left[\displaystyle\frac{\dot{x}^1_+(0)^2}{1+k\,u_+}-\displaystyle\frac{\dot{x}^2_+(0)^2}{1-k\,u_+}\right]+u_+\,\dot{v}_+(0)+u_-\,\dot{v}_-(0)+v_0\,,
\end{array}
\right. 
\label{diagc0geo}
\end{equation}
where the left and right velocities, $\dot{\bx}_{\pm}(0)$ and $\dot{v}_\pm(0)$, were carefully distinguished. 
 Note that rewriting the geodesic equation in BJR coordinates as 
\beq
\ddot{x}^1 + 2\dfrac {\dot{P}_{11}}{P_{11}}\, \dot{x}^1 = 0,
\quad
\ddot{x}^2 + 2\dfrac {\dot{P}_{22}}{P_{22}}\, \dot{x}^2 = 0,
\quad
\;\ddot{v} - P_{11}\dot{P}_{11}(\dot{x}^1)^2
-P_{22}\dot{P}_{22}(\dot{x}^2)^2
= 0 
\label{BJRPGeo}
\eeq
yields an easy check of (\ref{diagc0geo}).  The solution (\ref{diagc0geo}) is valid for $u < u_1=k^{-1}$ only; trajectories, depicted in dashed black lines, strongly diverge when $u\uparrow k^{-1} (=2)$ as can be seen in Fig.\ref{BJRxv}.
%%%%%%%%%%%%%%%%%%
%%%%%%%%%%%%%

\goodbreak
Comparison with (\ref{SB14}) shows that this solution
 corresponds  to (\ref{diagc0geo}) with $k=1$ and the $C^1$-assumption of \emph{unique} initial velocities,  
\beq
\dot\bx_+(0)=\dot\bx_-(0)=\dot\bx_0,
\qquad
\dot{v}_+(0)=\dot{v}_-(0)=\dot{v}_0.
\label{Steinbauer}
\eeq
Moreover, choosing  an appropriate initial condition at $u_i\ll0$ and letting $\lambda\to\infty$ in the Gaussian, the \emph{Steinbauer solution} (\ref{SB14}) is obtained \emph{for all $u$}, as shown in Fig.\ref{BJRxv}. The velocity $\dot{\bx}_0$ is the \emph{common left and right-side limit}. 
 We propose therefore to drop the $C^1$ property of transverse-space trajectories and use  (\ref{diagc0geo}) 
in the before and in the after-zone separately,  with \emph{different} initial velocities at $u=0\pm$,  namely
\beq
\dot{\bx}_0=\left\{\barraynb{ccl}
0 &\quad\text{in the before-zone}\; &u\leq0
\\
-c_0\bx_0&\quad\text{in the after-zone}\; &u>0
\earraynb\right.
\label{SBmatch}
\eeq
dictated by the ``jumping conditions" (\ref{dotx0+}) and
yielding our desired solution -- which, however, can \emph{not} be obtained by shrinking  Gaussian profiles.

\goodbreak

%%%%%%%%%%%%%%%%%%%
\section{Conclusion}
%%%%%%%%%%%%%%%%%%%

%\begin{redtext}

Using the ``hatted'' S-coordinates (\ref{Scoords}) is particularly convenient to determine the geodesics, since the latter are  simple straight lines in Minkowski space; the nontrivial behavior is hidden in the transformation formula (\ref{Phxx}).
In flat zones the B and S-coordinates coincide   and therefore the only problem is  matching correctly the geodesics -- allowing us to avoid ill-posed multiplication of distributional functions. This matching is analogous to  similar problems in continuous mechanics, and can be achieved by considering  the ``jump structure'' using the waves with smooth (Gaussian) profile with appropriate parameters, close to the distributional limit. 

This scheme leads to very short, clear and physically justified form of GW memory effects. 

Both in the flat before and after zones the motion is thus \emph{along straight lines with constant velocity}. This should not come as a surprise when remembering that a 4D GW spacetime can also be seen as the ``Bargmann description'' of a non-relativistic system  in $2+1$ dimensions \cite{Eisenhart,Bargmann}. For a sandwich wave the flat before and after zones describe a \emph{free non-relativistic particle} in transverse space. Our finding is therefore \dots a confirmation of Newton's First law \footnote{In contrast, Kulczycki and Malec  \cite{Malec} found circular trajectories in some Friedmann-Lema\^\i tre-Robertson-Walker background.}.
The $\delta$-function profile causes  jumps and breakings which correspond to the \emph{work}  done by the wave on the particle.

The impulsive wave \emph{metric} is obtained by shrinking Gaussians. However, the \emph{coordinate-change formula} (\ref{BBJRtrans}) between B and BJR coordinates \emph{fails} for geodesics. Solving the geodesic equations separately in B and in BJR coordinates, (\ref{BX}) and (\ref{FreeFallinGW}), respectively, their comparison yields instead (\ref{goodXx}). 
 
By shrinking Gaussians, we get smooth transverse trajectories which match our exact solution (\ref{diagc0geo}) in the after-zone $u>0$ but \emph{not} in the before-zone, see Fig.\ref{BJRxv}; they yield instead, for \emph{all} $u$, the smooth Steinbauer solutions  (\ref{SB14}) with no velocity jump. However the velocity jump  (\ref{Deltadotv}) is \emph{mandatory}; it can be taken into account by working in the flat zones separately and then gluing the solutions respecting the jump conditions (\ref{Pjump}), (\ref{percussion}) resp. (\ref{BXveljump}) and (\ref{Vdiff}).

The most dramatic effect, which  arises only in the impulsive case and is absent for smooth profile, is the \emph{discontinuity} suffered by the lightlike coordinate $V(U)$.\footnote{In our previous papers the behavior of $V(0)$ was neglected by the practical reason  that it is \emph{quadratic} and therefore irrelevant for eventual observations of GWs: even the first-order velocity effect is very small \cite{Lasenby}.}
Our  (\ref{Vdiff}) is indeed consistent with (\ref{Vjump}) when we remember that the BJR coordinate $v(u)$ \emph{is} continuous, as noted before.  Alternatively, the  B-trajectories are the images of those trivial straight ones in Minkowski space, given in (\ref{Xxhx}). 

One may protest that discontinuous worldlines are unphysical. We agree however we also argue  that approximating real smooth wave profiles (as Gaussians with ``width'' $\lambda^{-1}$) is a mathematical idealization in itself. Those continuous trajectories do exhibit, as illustrated by figs.\ref{B3geo} and 
\ref{squeezegeoB}, sharp increases of the $V$ coordinate, which do tend to (\ref{Vdiff}) when $\lambda\to\infty$ \footnote{
  Podolsk\'y and Ortaggio \cite{Podolsky} had also found discontinuous trajectories in (anti-)de Sitter space-time.}.

The situation is reminiscent of that \emph{geometrical optics}~:  approximating the real situation with a sharp change of the refractive index leads to the Snell law and to the recently proposed Spin Hall Effect of Light \cite{SHEL} which are  ``unphysical idealizations'', --- which yield, nevertheless, predictions in  agreement with observations.  
In conclusion, we believe that the jump \emph{is} physical --- even if it is very small: let alone the millions-of-kms-long arms of future LISA might be too short when compared to the distances from the sources. 

%\end{redtext} 

\goodbreak

%\%\%\%
\begin{acknowledgments} 
We are grateful to Gary Gibbons for his interest and collaboration at the early stages of this project, and for his continuous advice. Correspondence with Edward Malec and Marcello Ortaggio is acknowledged.
PH thanks  the \emph{Institute of Modern Physics} of the Chinese Academy of Sciences in Lanzhou  for hospitality where part of this research was conducted. Support by the National Natural Science Foundation of China (Grant No. 11575254) is acknowledged.
\end{acknowledgments}
\goodbreak

%%%%%%%%%%%%%%%%%%%%%%%%%%%%%%%%%%%%%%%%%%%%%%%%%%%%%%%%%%%%%%%%%%%%%%%%%%%%%%
%%%%%%%%%%%%%%%%%%%%%%%%%%%%%%%%%%%%%%%%%%%%%%%%%%%%%%%%%%%%%%%%%%%%%%%%%%%%%%

\end{document}